\documentclass[11pt,preprint]{aastex}

\newcommand{\FUSE}{{\it FUSE}}
\newcommand{\FUSElong}{{\it Far Ultraviolet Spectroscopic Explorer}}
\newcommand{\IRAS}{{\it IRAS}}

\newcommand{\kms}{km~s{$^{-1}$}}

\newcommand\etal{et~al.}
\newcommand{\cd}{cm$^{-2}$}

\newcommand{\Htwo}{H$_2$}
\newcommand{\NHtwo}{N$_{\rm H2}$}
\newcommand{\NHI}{N$_{\rm HI}$}

\begin{document}

\title{MOLECULAR HYDROGEN IN INFRARED CIRRUS } 

\author{Kristen Gillmon\altaffilmark{1} \& J. Michael Shull}

\affil{University of Colorado, Department of Astrophysical \&
     Planetary Sciences, \\ CASA, 389-UCB, Boulder, CO 80309 }  

\altaffiltext{1}{Now at Department of Astronomy, 
  University of California, Berkeley, CA 94720 }  

\email{kristen.gillmon@colorado.edu, mshull@casa.colorado.edu}

\begin{abstract}

We combine data from our recent \FUSE\ survey of interstellar molecular 
hydrogen absorption toward 50 high-latitude AGN with COBE-corrected
\IRAS\ 100 $\mu$m emission maps to study the correlation of infrared 
cirrus with \Htwo.  A plot of the \Htwo\ column density vs.\ 
IR cirrus intensity shows the same transition in molecular fraction,
$f_{\rm H2}$, as seen with total hydrogen column density, N$_{\rm H}$. 
This transition is usually attributed to \Htwo\ ``self-shielding", and 
it suggests that many diffuse cirrus clouds contain \Htwo\ in significant 
fractions, $f_{\rm H2} \approx$ 1--30\%.  These clouds cover $\sim50$\%
of the northern sky at $b > 30^{\circ}$, at temperature-corrected
100 $\mu$m intensities $D_{100}^{(T)} \geq 1.5$ MJy~sr$^{-1}$.     
The sheetlike cirrus clouds, with hydrogen densities 
$n_H \geq 30$ cm$^{-3}$, may be compressed by dynamical processes 
at the disk-halo interface, and they are conducive to \Htwo\ formation on
grain surfaces.  Exploiting the correlation between \NHtwo\ and 100~$\mu$m 
intensity, we estimate that cirrus clouds at $b > 30^{\circ}$
contain $\sim 3000~M_{\odot}$ in \Htwo.  Extrapolated over the inner 
Milky Way, the cirrus may contain $10^7~M_{\odot}$ of \Htwo\ and
$10^8~M_{\odot}$ in total gas mass.  If elevated to 100 pc, their
gravitational potential energy is $\sim10^{53}$ erg.

\end{abstract}


\section{INTRODUCTION}

In an attempt to understand the molecular content and physical
characteristics of interstellar gas in the low Galactic halo, 
we exploit infrared and ultraviolet data from two NASA satellites:   
the {\it Infrared Astronomical Satellite} (\IRAS) mission of
1983 and the \FUSElong\ (\FUSE) satellite of 1999--2005.  The 
combination of infrared emission and ultraviolet absorption along 
sight lines to 45 active galactic nuclei (AGN) allows us to 
correlate the infrared cirrus emission intensity with the
molecular hydrogen (\Htwo) absorption column density at select 
locations at high Galactic latitude.    

\IRAS\ mapped the sky in four infrared bands centered on 12, 25, 60, 
and 100~$\mu$m.  Low \etal\ (1984) introduced one of the most surprising 
results from the \IRAS\ maps: diffuse filamentary dust clouds 
that pervade our Galaxy at high latitudes, even in the direction of 
the Galactic poles.  These ``infrared cirrus" clouds are most prominent 
at long wavelengths, 100~$\mu$m, but some can be seen in the 
60~$\mu$m, 25~$\mu$m, and 12~$\mu$m bands.  Comparisons between 
\IRAS\ dust maps and maps of 21~cm emission reveal a generally good 
correlation between neutral hydrogen (Hartmann \& Burton 1997) 
and cirrus dust filaments (Figure 1).  
Because molecular hydrogen (\Htwo) forms catalytically on the surface 
of dust grains (Hollenbach, Werner, \& Salpeter 1971), with significant 
formation rates for grain temperatures $T_{\rm gr} < 100$~K and gas 
temperatures $T_{\rm gas} < 300$~K (Hollenbach \& McKee 1979; Shull 
\& Beckwith 1982), the cold, dusty conditions of the infrared cirrus 
clouds are favorable for the formation of H$_2$.  Thus, it is plausible 
that some fraction of the hydrogen atoms in the cirrus clouds are bound 
into molecules.  

Previously, the presence of \Htwo\ in infrared cirrus was inferred 
indirectly.  First, under the assumption that the infrared 
emission and total hydrogen column density, N$_{\rm H}$, are proportional, 
regions of high dust/\ion{H}{1} ratio, termed ``infrared excess", were 
attributed to the presence of \Htwo\ (de~Vries, Heithausen, \& Thaddeus 
1987; Desert, Bazell, \& Boulanger 1988; Reach, Koo, \& Heiles 1994; 
Moritz \etal\ 1998; Schlegel, Finkbeiner, \& Davis 1998, hereafter 
SFD98).  Second, the detection of CO in dense cirrus clouds suggests 
that the diffuse cirrus clouds should contain \Htwo\ as well.  
Weiland \etal\ (1986) compared CO maps from Magnani, Blitz, \& Mundy (1985) 
to \IRAS\ maps of infrared cirrus.  Each of the 33 CO clouds had a cirrus 
counterpart with similar 
morphology.  This work established that at least some of the infrared 
cirrus cloud cores contains CO gas.

Unfortunately, there is currently no experiment that can map diffuse 
\Htwo\ emission, either in the 2.12~$\mu$m [(1--0) S(1)] vibrational 
line or in the S(0), S(1), S(2) pure rotational lines at 
28~$\mu$m, 17~$\mu$m, and 12~$\mu$m, respectively.  
Although \Htwo\ is over $10^4$ times more abundant than CO, the 
ultraviolet and infrared fluorescent emission of \Htwo\ is very weak.  
Ultraviolet absorption-line spectroscopy is therefore the primary means 
for detecting cold \Htwo\ in diffuse clouds.  However, it 
requires background sources with sufficient UV flux to provide 
adequate signal-to-noise ratio (S/N) to detect the weak \Htwo\ lines.  
The first major project to conduct 
such observations was the {\it Copernicus} mission of the 1970s 
(Spitzer \& Jenkins 1975).  However, its sensitivity limited the possible 
background sources to early-type stars within about 500~pc of the 
Sun.  Most OB stars that fit this criterion are at low Galactic latitude, 
and they suffer from confusion and dust extinction in the Galactic plane.  
Individual features in the infrared cirrus cannot be discerned at low 
Galactic latitudes, and these stellar sight lines are not effective probes 
of the dusty filaments.  

The \FUSE\ satellite, which has been observing the ultraviolet sky 
since 1999, has expanded the opportunities for detecting \Htwo.
The increased sensitivity of \FUSE\ ($m_{\rm{lim}} \approx 15.5$\,mag) 
over {\it Copernicus} ($m_{\rm{lim}} \approx 8$\,mag) allows us to use 
more distant stars as well as active galactic nuclei (AGN) as background 
sources.  
Our \FUSE\ survey of \Htwo\ toward high-latitude AGN (Gillmon \etal\
2005) is particularly well suited for probing \Htwo\ in infrared cirrus.  
The high-latitude sight lines avoid the confusion of the 
Galactic disk, and they provide long path lengths through the 
Galactic halo.  In addition, the random distribution of AGN on the sky 
samples a range of infrared cirrus emission intensities. 
The main limitation of using ``pencil-beam" (absorption)
sight lines to detect \Htwo\ in infrared cirrus is the 
inability to determine whether the gas and dust detected along a given 
sight line are physically associated. Therefore, we must rely on
indirect correlations between cirrus and \Htwo\ absorption. 

In this paper, we compare the \Htwo\ column densities in the survey 
by Gillmon \etal\ (2005) with the corresponding infrared cirrus 
fluxes (SFD98).  By establishing a correlation between the two, we 
assert that at least some of the detected \Htwo\ resides in the cirrus 
clouds.  In \S~2 we describe the data acquisition and analysis for both 
\IRAS\ and \FUSE.  In \S~3 we compare the cirrus emission and \Htwo\ 
absorption and discuss the correlation of the two.  Exploiting 
this correlation and summing over the distribution of \Htwo\ column 
densities with 100 $\mu$m cirrus intensity, we estimate the total 
\Htwo\ mass ($\sim 10^7~M_{\odot}$) in cirrus clouds around the Milky Way. 
With molecular fractions ranging from 1--30\%, the total cirrus 
mass throughout the Milky Way is $\sim 10^8~M_{\odot}$.  
In \S~4 we summarize our results and the implications of finding 
this amount of gas in the low halo of the Milky Way.

\section{DATA ACQUISITION AND ANALYSIS}

\subsection{\FUSE\ Observations}

The \Htwo\ absorption data were taken from \FUSE\ spectra
of AGN, using standard data analysis techniques 
(Tumlinson \etal\ 2002; Gillmon \etal\ 2005).   
Studies of \Htwo\ have been a major part of the \FUSE\ science plan.  
The satellite, its mission, and its on-orbit performance are 
described in Moos \etal\ (2000) and Sahnow \etal\ (2000).
Scientific results on interstellar \Htwo\ have appeared in a number 
of papers (Shull \etal\ 2000; Snow \etal\ 2000; Rachford \etal\ 2002;
Richter \etal\ 2001, 2003; Tumlinson \etal\ 2002; Shull \etal\ 2005).
The resolution of \FUSE\ varies from $R = \lambda / \Delta \lambda
= 15,000 - 20,000$ across the far-UV band. 
All observations were obtained in time-tag (TTAG) mode,
using the $30'' \times 30''$ LWRS aperture, with  
resolution $\sim20$ \kms\ at 1050 \AA. The S/N of the co-added 
data ranges from 2--11 per pixel; the S/N per resolution element 
varies with spectral resolution, which is not fixed in our survey.  
Most of the data were binned by 4 pixels before analysis, with
the rare case of binning by 2 or 8 pixels.

The data in this paper were taken from our high-latitude \Htwo\ 
survey (Gillmon \etal\ 2005), which describes our search for \Htwo\ 
absorption along 45 sight lines to background AGN at Galactic latitudes 
$|b| > 20^{\circ}$.  The 45 AGN in the survey by Gillmon \etal\ (2005) 
were a subset of the 219 {\em FUSE} targets selected in Wakker
\etal\ (2003) as candidates for the analysis of Galactic O\,VI. 
These targets probe diffuse gas in both the local 
Galactic disk and low Galactic halo.  Of the available AGN at high 
latitude, 45 sight lines were chosen, based on an imposed S/N requirement 
of $\rm{(S/N)_{bin}>4}$ or $\rm{(S/N)_{pix}>2}$ with 4-pixel binning.  
The observed \Htwo\ lines arise from the Lyman and Werner electronic 
transitions, from the ground electronic state, X~$^{1}\Sigma^{+}_{g}$, 
to the excited states, B~$^{1}\Sigma_{u}^{+}$ (Lyman bands) and
C~$^{1}\Pi_u$ (Werner bands). The rotational-vibrational lines arise 
from the ground vibrational state and a range of rotational states.  
Our analysis was restricted to ten vibrational-rotational bands, 
Lyman (0--0) to (8--0) and Werner (0--0), which are located between 
1000~$\rm{\AA}$ and 1126~$\rm{\AA}$.  The vibrational state notation 
is ($v_{\rm{upper}}-v_{\rm{lower}}$). In most sight lines, we observed 
absorption lines from rotational states $J=0$--3, and sometimes up 
to $J=4$. 

The end product of the \Htwo\ absorption-line analysis 
is the column density, \NHtwo\  (cm$^{-2}$), the physical density of 
\Htwo\ molecules, integrated along the sight line.  Each absorption line 
was fitted with a Voigt profile in order to determine the equivalent width, 
a measure of the absorbed light in the line.  The equivalent widths were 
fitted to a curve of growth, to find the column density, N$_{\rm H2}(J)$
in each rotational state $J$.  The points for each $J$ were tied together 
during the fitting to produce a single, consistent column density for 
each rotational state, N($J$).  The sum of the column densities 
in all rotational states then gives the total column density, \NHtwo. 
FUSE spectra of 87\% (39 of 45) of the observed AGN showed detectable 
\Htwo\ absorption, with column densities ranging from 
\NHtwo\ $= 10^{14.17-19.82}$~\cd.  The \FUSE\ survey is sensitive to 
\NHtwo\ $>10^{13.8-14.6}$~\cd,  depending on the S/N (2--11 per 
pixel) and spectral resolution ($R=15,000$--20,000).  

\subsection{IRAS Dust Maps}

To obtain infrared cirrus emission intensities (MJy~sr$^{-1}$), we use 
the 100~$\mu$m maps presented by SFD98.  These were a composite of data 
from the \IRAS\ mission of 1983 and the {\it COBE} mission of 1989--1990, 
capitalizing on the strengths of each.  
Because the interstellar dust emits like a ``grey body", the emission
intensity is sensitive to the dust temperature.  As a result,
two regions with the same dust column density but different dust temperatures
will have different infrared intensities.  To correct for this effect, SFD98
used the ratio of the 100~$\mu$m and 240~$\mu$m {\it COBE} maps to produce
a map of dust color temperature.  They used this ratio to correct the
100~$\mu$m \IRAS\ map so that it is proportional to dust column density.
\IRAS\ mapped the sky in four 
broadband infrared channels, centered at 12, 25, 60, and 100 $\mu$m 
with a resolution of $\sim5'$, while {\it COBE} mapped the sky in 10 broad 
photometric bands from 1 to 240 $\mu$m at a resolution of $\sim0.7^\circ$.  
Before combining the data sets, SFD98 took great care in the difficult 
removal of the zodiacal foreground emission and \IRAS\ striping artifacts 
that arise from differences in solar elongation between scans.  
Confirmed point sources were also removed.  The maps were combined in such 
a way as to preserve the {\it COBE} calibration and \IRAS\ resolution.  

Maps of the temperature-corrected 100~$\mu$m intensity, $D_{100}^{(T)}$, 
are presented for the northern Galactic hemisphere (Figure 2) and for the 
southern Galactic hemisphere (Figure 3).  We overplot the locations of 
the 45 sight lines from the \FUSE\ \Htwo\ survey:  
28 northern AGN and 17 southern AGN. 

\section{Comparisons of \Htwo\ and Infrared Cirrus}

\subsection{The \Htwo\ Self-Shielding Transition in N$_H$}

Before the \FUSE\ mission, the direct detection of \Htwo\ in infrared 
cirrus by UV absorption-line spectroscopy was prevented mainly by a lack 
of background sources at high Galactic latitude.  Even though this problem 
has been alleviated by \FUSE's  ability to observe bright AGN as background 
sources, another problem with absorption-line spectroscopy 
along ``pencil-beam" sight lines comes to the forefront.  Absorption studies 
lack morphological information and cannot provide distances to clouds along 
the line of sight.  Therefore, it is difficult to identify the 
gas that gives rise to the detected column density.  The infrared cirrus 
problem is a classic case.  For any given sight line, it is possible that 
the \Htwo\ absorption is not physically associated with the infrared cirrus 
along the beam.  If this is the case, then comparing \NHtwo\ with the 
cirrus dust column density would lead to erroneous results.  It would 
be helpful to determine whether the diffuse \Htwo\ along all 
sight lines resides in the cirrus.  If no other component of the 
diffuse ISM harbored significant amounts of \Htwo, then \NHtwo\ for a 
given sight line could safely be associated with other cirrus properties.  

In this section, we investigate this possibility, based on a property of 
\Htwo\ called ``self-shielding".  Following line absorption of UV photons 
from the mean interstellar radiation field, \Htwo\ decays approximately 
11\% of the time to the dissociative continuum.  
As a result, the molecular fraction,
\begin{equation}
    f_{\rm H2} = \frac { 2 {\rm N}_{\rm H2} } 
          { {\rm N}_{\rm HI} + 2 {\rm N}_{\rm H2} }  
       \equiv \frac {2 N_{\rm H2}} {{\rm N}_{\rm H}}  \; ,
\end{equation}
is generally larger in clouds with a greater total hydrogen column 
density, N$_{\rm H}$ = \NHI\ + 2\NHtwo.  Molecules on the outside of the 
cloud shield those in the interior from dissociating UV 
(Hollenbach, Werner, \& Salpeter 1971; Black \& Dalgarno 1976;
Browning, Tumlinson, \& Shull 2003). 

In optically thin clouds, the density of molecules can be
approximated by the equilibrium between formation and destruction,
\begin{equation}
 f_{\rm H2} \approx \frac {2 n_H R(T_{\rm gas}, T_{\rm gr}, Z) }
               {\beta \; \langle f_{\rm diss} \rangle }
     \approx (10^{-5}) R_{-17} \; n_{30} \left(
         \frac {\beta_0} {\beta} \right)  \; .
\end{equation}
In this formula, the numerical value for $f_{\rm H2}$ is scaled to
fiducial values of hydrogen density, $n_H$ (30 cm$^{-3}$), \Htwo\
formation rate coefficient, $R$ ($10^{-17}$ cm$^3$~s$^{-1}$), and
mean \Htwo\ pumping rate in the FUV Lyman and Werner bands,
$\beta_0 =  5 \times 10^{-10}$ s$^{-1}$. The \Htwo\ photodissociation
rate is written as $\langle f_{\rm diss} \rangle \beta$, where the
coefficient $\langle f_{\rm diss} \rangle \approx 0.11$ is the
average fraction of FUV excitations of \Htwo\ that result in decays
to the dissociating continuum.  The \Htwo\ formation rate per unit
volume is written as $n_H n_{\rm HI} R$, where the coefficient $R$
depends on the gas temperature, grain surface temperature, and gas
metallicity ($Z$). The metallicity dependence comes from the assumed
scaling of grain-surface catalysis sites with the grain/gas ratio.
For sight lines in the local Galactic disk, this rate coefficient has
been estimated (Jura 1974) to range from
$R = (1-3) \times 10^{-17}$ cm$^3$~s$^{-1}$ at solar metallicity.
This standard value for $R$ is expected to apply at suitably
low temperatures of gas ($T_{\rm gas} \leq 300$~K) and grains
($T_{\rm gr} \leq 100$~K) as discussed by Shull \& Beckwith (1982)
and Hollenbach \& McKee (1979).

The effects of self-shielding were observed by plotting $f_{\rm H2}$ 
versus N$_{\rm H}$, the total column density of hydrogen. As \Htwo\ 
absorption lines in the Lyman and Werner bands 
become optically thick, the rate ($\beta$) of UV pumping and molecular 
dissociation diminish.  In this way, the presence of \Htwo\ 
screens molecules in the inner portions of the cloud from dissociation.  
A transition from low 
($f_{\rm{H2}} \approx 10^{-5}$) to high ($f_{\rm{H2}} > 10^{-2}$) molecular 
fractions at N$_{\rm{H}} \ge 5\times 10^{20}$~cm$^{-2}$ was noted in the 
{\it Copernicus} \Htwo\ survey (Savage \etal\ 1977).  A recent 
\FUSE\ survey of Galactic disk stars (Shull \etal\ 2005)
provided similar results (Figure 4).  
 
In assessing the molecular content of the cirrus, we consider three
generic cases.  The first possibility (Case~I) is that all the 
detected H$_2$ resides in cirrus clouds, with no contribution from 
foreground clouds.  Under the assumption that the total hydrogen density, 
$n_{\rm{H}}$, is proportional to the number density of dust grains in any 
given interstellar cloud (Hollenbach \& McKee 1979), the cirrus dust 
column density should be proportional to the amount of  
N$_{\rm{H}}$ associated with \Htwo.  
A plot of \NHtwo\ vs.\ dust column density should then show the familiar 
self-shielding transition of molecular hydrogen.  In Case~II,
some of the observed \Htwo\ lies in the cirrus, but some resides in 
another component of the ISM not visible in the cirrus maps.  
A plot of \NHtwo\ vs.\ dust column density would show some points that 
follow the self-shielding transition (those corresponding to 
cirrus).  However, the transition would be broadened by points that are 
randomly distributed.  If a significant portion of the detected \Htwo\
along a sight line is not in cirrus, \NHtwo\ will not correlate with 
dust column density.  This effect would produce sight lines with low dust 
column density and significant amounts of \Htwo.  The extreme situation
is Case~III, in which none of the \Htwo\ resides in cirrus clouds.  
We consider this unlikely, given the number of observed regions with 
``infrared excess" (high dust/\NHI\ ratios) and the detection of CO in 
denser cirrus clouds.  Once the \NHtwo\ in cirrus exceeds 
$10^{14.5}$~cm$^{-2}$, it should be detectable by \FUSE.  
For Case~III to be consistent with our survey, the covering fraction 
of cirrus regions with detectable \Htwo\ would need to be quite small, 
whereas the cirrus covering factor is observed to be  
$\sim50$\% at $b > 30^{\circ}$ (see \S~3.4). 

The issue then comes down to whether Case I or Case II is a better
description of the cirrus-\Htwo\ correlation.  
Figure 5 presents a plot of \NHtwo\ along sight lines to the 45 AGN 
in the \FUSE\ survey vs.~the temperature-corrected IR flux at each 
location from the SFD98 maps (blue diamonds).  If the 
temperature-corrected flux is proportional to dust column density, 
this plot can be used to test the three cases mentioned above.  The blue 
diamonds appear to show the self-shielding transition discussed in Case I, 
implying that much of the detected \Htwo\ is in the cirrus.  

\subsection{AGN Behind Regions of Low Cirrus}

In this section, we describe five additional sight lines shown in red 
(Figure 5) toward the regions of lowest cirrus.  These additional sight
lines were chosen to explore whether the correlation between \NHtwo\
and 100~$\mu$m cirrus is universal.   
Table 5 of SFD98 lists the coordinates of the regions of lowest 
(temperature-corrected) 100 $\mu$m intensity, $D_{100}^{(T)}$, from their 
maps.  We searched the \FUSE\ archive for targets within $5^{\circ}$ of the 
given coordinates and found five targets behind regions of low cirrus 
($D_{100}^{(T)}\le0.5$ MJy\,sr$^{-1}$) that also had \FUSE\ data with 
sufficient (S/N)$_{\rm{pix}}>2$ to conduct an \Htwo\ analysis.
These five targets are in addition to those in the Gillmon \etal\ (2005) 
sample of 45 AGN.  They are listed in Table 1 and shown as asterisks in 
Figure 5.  

Four of these five targets showed no evidence of H$_2$, with a typical 
upper limit of \NHtwo\ $\le 10^{14.5}$~\cd.  This result 
lends further credence to the theory that most of the observed \Htwo\ is 
in the cirrus (Case I).  However, one target, UGC\,5720, showed significant 
\Htwo\ absorption lines (Figure 6).  An analysis, as described in \S~2.1, 
yielded a significant column density \NHtwo\ $= 10^{18.79 \pm 0.05}$~\cd.  
In \S~3.3, we explore possible explanations for this anomalous sight line 
and for the spread in the self-shielding transition.

\subsection{The \Htwo\ Transition seen in Cirrus} 

Figure 5 plots \NHtwo\ vs.\ temperature-corrected intensity, 
$D_{100}^{(T)}$ (MJy\,sr$^{-1}$), and exhibits a clear self-shielding 
transition of H$_2$.  This indicates that 
a significant fraction of the detected \Htwo\ resides in the infrared 
cirrus.  However, with small-number statistics on 50 AGN sight lines, the 
transition is not sharp, occurring between log~$D_{100}^{(T)}$ = 0.2--0.5.
There is one outlying point (UGC\,5720) with low cirrus intensity, 
log~$D_{100}^{(T)} = -0.19$, but a significant column 
density, log~\NHtwo\ $= 18.79 \pm 0.05$.  This outlier suggests that 
not all the detected \Htwo\ resides in cirrus clouds, and that other 
components of the diffuse gas may harbor detectable amounts of molecules.  

In some sight lines that intercept non-cirrus clouds with low 
100~$\mu$m intensity, the gas may undergo an
early molecular transition and appear with higher \NHtwo.
There are other explanations that might produce
the same effects, even if all the H$_2$ resides in cirrus clouds.  
One consideration is how well the temperature-corrected flux map 
traces the dust column density.  To correct for temperature variations, 
SFD98 created a temperature map from the ratio of the {\em COBE} 
100~$\mu$m and 240~$\mu$m maps.  Thus, low-resolution 
(1.1$^{\circ}$) temperature maps were used to correct the $5'$ resolution 
\IRAS\ map to produce a 100~$\mu$m map at $6.'1$ resolution.  
It is likely that the dust temperature varies on 
smaller scales than can be resolved by this method, and a more accurate 
temperature correction might tighten the self-shielding transition.  It is 
also possible that UGC\,5720 lies behind a region of cold cirrus dust 
unresolved by the temperature map. Thus, a significant portion of 
the dust column density might not be indicated by the observed flux.  

A related consideration is that the \IRAS\ map might not resolve 
significant variations in dust column density.  The sight lines observed 
by \FUSE\ absorption probe gas along a very narrow beam, while the 
\IRAS\ beam is considerably larger.  Thus, if there were significant 
structure on smaller scales than the \IRAS\ beam, \FUSE\ could 
observe a denser clump of \Htwo\ while the \IRAS\ emission would be 
``beam-diluted".  A higher resolution infrared map might tighten the 
self-shielding transition and/or 
bring the outlier UGC\,5720 onto the correlation.  Figure 7 shows a 
$2^{\circ}\times2^{\circ}$ section of the SFD98 flux map centered 
on UGC\,5720.  Large fluctuations in temperature-corrected flux near the 
position of the AGN suggest that the sight line may be picking up a clump 
of cirrus with higher dust column density unresolved by these maps.

Finally, there is the possibility that there are multiple cirrus 
features superimposed along a line of sight.  Duel \& Burton (1990) compared 
the morphology of cirrus clouds and H~I maps in various velocity intervals 
to show that cirrus features that appear simple are, in some cases, 
superpositions of kinematically distinct components.  This idea of a
``concatenation of clouds" was proposed for translucent \Htwo\ clouds 
(Browning \etal\ 2003) to explain the high levels of
\Htwo\ rotational excitation in systems with large \NHtwo. 
These authors suggested that, if a sight line intercepts multiple, 
physically distinct cloud components, the \Htwo\ will be exposed to a 
radiation field enhanced over that expected for a single, contiguous cloud
with the same total column density.  The enhanced \Htwo\ destruction rate 
from the stronger UV radiation would reduce the mean molecular 
fraction and produce a gradual transition to higher \NHtwo.  Some 
sight lines with seemingly high dust column density, 
but low \NHtwo, may actually be probing multiple, superimposed 
filaments with high integrated dust column density.

\subsection{Mass of the Cirrus Clouds}

With the empirical correlation (Figure 5) between IR cirrus and \Htwo\ 
column density, we can make a quantitative estimate of the \Htwo\ mass 
in the diffuse cirrus clouds.   We begin with the \IRAS\ maps of the 
northern Galactic hemisphere.  We assume that these cirrus clouds lie 
at elevation $z \approx (100~{\rm pc})z_{100}$ above 
the Milky Way plane, and that clouds of intensity $D_i$ cover a fraction 
$f_c(D_i)$ of the planar area at $b \geq 30^{\circ}$.  The total 
\Htwo\ mass in this planar cloud deck is then,
\begin{equation} 
   M_{\rm H2} = (\pi R^2) (2 m_{\rm H2}) 
         \sum_i {\rm N}_{\rm H2}(D_i) f_c(D_i) \Delta D_i  \approx 
     (2400~M_{\odot}) \left[ \frac {\langle {\rm N}_{\rm H2} \rangle }
        {10^{18.5}~{\rm cm}^{-2}} \right] \left[ \frac {\langle f_c \rangle} 
        {0.5} \right] z_{100}^2   \; , 
\end{equation}
where \NHtwo $(D_i)$ is the mean \Htwo\ column density corresponding
to cirrus intensity $D_i$ (Figure 5) and where $R = z/\tan(b) 
\approx (173~{\rm pc})z_{100}$ is the radius of the cirrus disk, at 
elevation $z$ subtended by the cone at $b = 30^{\circ}$.  For this 
estimate, we have assumed that the cirrus covers a fraction 
$f_c \approx 0.5$ of the sky at $b \geq 30^{\circ}$, independent of 
intensity.  

We now make a more careful calculation, summing over the actual data
from \IRAS\ and \FUSE.  Figure 8 shows the 
distribution of cirrus covering factors, $f_c(D_i)$, averaged over 
northern hemisphere regions with $b$ greater than
$30^{\circ}$, $40^{\circ}$, and $50^{\circ}$, respectively.  
For our fiducial cone at
$b \geq 30^{\circ}$, approximately 50\% of the sky is covered
by cirrus with intensity log~$D_{100}^{(T)} \geq 0.2$, 
the value corresponding to the \Htwo\ transition (Figure 5).    
Performing the full summation (eq.\ 3) over 8 logarithmic intensity 
bins above the transition, of width $\Delta$(log~$D_i$) = 0.1, between 
log~$D_i = 0.2$ and 1.0, we find a total \Htwo\ mass of 
$(2600~M_{\odot}) z_{100}^2$ contained in the cirrus at $b \geq 30^{\circ}$.

To convert this calculation to the inner Milky Way, within the solar
circle, we multiply by a factor $2$, for the northern and southern Galactic 
hemispheres, and scale by a factor $(8.2~{\rm kpc}/0.173~{\rm kpc})^2 \approx 
2250 z_{100}^{-2}$ to account for the number of similar conical areas
around the Galactic disk.  Note that this total \Htwo\ mass is independent 
of the assumed cirrus elevation, $z$, since the area-scaling
cancels the factor $z_{100}^2$ in equation (3).  We arrive at an 
extrapolated total molecular mass over the inner Milky Way,
$M_{\rm H2}^{\rm cirrus} \approx 10^7~M_{\odot}$,
assuming that the cirrus along the AGN sightlines at $b > 30^{\circ}$ 
is typical.  The molecular fractions of these cirrus clouds, with
\NHtwo\ $\geq 10^{18.5}$~\cd, range from 1--30\% for 
log~$D_{100}^{(T)}$ = 0.2--0.5 (see Figure 6 of 
Gillmon \etal\ 2005), with an average $\langle f_{\rm H2} \rangle = 0.1$
for clouds with log~\NHtwo\ $\approx 18.5$.  Therefore, we estimate the 
total gas mass in the cirrus to be $\sim 10^8~ M_{\odot}$.  

The characteristics of the cirrus clouds along our AGN sight lines 
can be verified by computing the ``dust-to-gas" ratio, defined as 
the ratio of 100~$\mu$m cirrus intensity, $D_{100}^{(T)}$ (MJy~sr$^{-1}$),
to H~I column density, \NHI\ (cm$^{-2}$).  Table 2 gives these values 
and their ratio for 16 of our AGN sight lines observed in 21~cm 
(Lockman \& Savage 1995).  This ratio ranges from $0.66 \times 10^{-20}$ 
MJy~sr$^{-1}$~cm$^2$ toward PG~0953+414 to $4.3 \times 10^{-20}$ 
MJy~sr$^{-1}$~cm$^2$ toward 3C~273. The mean value and standard deviation 
are $(1.43 \pm 0.41) \times 10^{-20}$ MJy~sr$^{-1}$~cm$^2$, when we exclude 
the anomalous sight line to 3C~273, which lies behind Radio Loop I and 
the North Polar Spur.  This mean ratio is in excellent agreement with 
the mean value, $(1.4 \pm 0.3) \times 10^{-20}$ MJy~sr$^{-1}$~cm$^2$,
found by Boulanger, Baud, \& van Albada (1985) in a 
$20^{\circ} \times 18^{\circ}$ field at high Galactic latitude.

\section{SUMMARY AND FUTURE WORK}

We have undertaken a comparison between the column density, \NHtwo, and
the 100~$\mu$m cirrus intensity for a total of 50 sight lines.  For the 
cirrus, we used the temperature-corrected maps of Schlegel, Finkbeiner,
\& Davis (1998), and adopted \Htwo\ column densities from our 
high-latitude survey (Gillmon \etal\ 2005). 
The presence of a clear correlation between UV (\Htwo) absorption and 
IR (cirrus) emission indicates that a significant fraction of the \Htwo\ 
is physically associated with the cirrus clouds.  However, the 
self-shielding transition of \Htwo\ fraction used to define the 
correlation is not sharp. The existence of one outlying sight line 
suggests either that some of the detected H$_2$ may exist in another component 
of the diffuse ISM, or that the limited resolution of the infrared maps 
is obscuring the physical conditions.  Of the three possible cases for 
\Htwo--cirrus connections laid out in \S~3.1, Case I or II best describe 
the data.

Put simply, \Htwo\ is contained in most, if not all diffuse cirrus clouds.  
At Galactic latitudes $b > 30^{\circ}$, approximately 50\% of the sky 
is covered with cirrus, at temperature-corrected 100 $\mu$m intensities 
$D_{100}^{(T)} \geq 1.5$ MJy~sr$^{-1}$.  With this correlation, we have 
found a convenient means of identifying the best extragalactic sight 
lines for ``\Htwo-clean" far-UV absorption studies of intergalactic 
or interstellar matter.  Conversely, if the goal is to study \Htwo\ 
at the disk-halo interface, the cirrus maps would be a good guide. 

We also made a rough estimate of the \Htwo\ mass contained in
these cirrus clouds. Exploiting the \Htwo--cirrus correlation,
we summed the distributions of IR cirrus intensity and \Htwo\ 
column density to find $\sim 10^7~M_{\odot}$ in cirrus \Htwo\ and 
$\sim 10^8~M_{\odot}$ in total hydrogen, distributed over the 
Milky Way disk-halo interface, within the solar circle.  Above the 
self-shielding transition, these diffuse halo clouds have molecular 
fractions ranging from 1--30\%, for column densities 
N$_{\rm H} \geq 10^{20.4}$~cm$^{-2}$ and 
\NHtwo\ $\geq 10^{18.5}$~cm$^{-2}$ (Gillmon \etal\ 2005).  
To support such high molecular fractions by \Htwo\ formation 
on grain surfaces, the cirrus clouds are probably compressed sheets 
with densities $n_H \geq 30$ cm$^{-3}$, in which the gas and grains 
remain sufficiently cold to form \Htwo. On average,  
the IR cirrus clouds may actually have higher molecular 
fractions, on average, than diffuse clouds in the Galactic disk.   
This point was also made by Reach \etal\ (1994), who estimated
that the \Htwo/H~I transition in denser cirrus clouds occurs at 
N$_{\rm H} \approx 4 \times 10^{20}$ \cd, on the basis of fits
to the far-infrared excess.  This transition column density is 
almost twice that found here, probably because the
infrared-excess technique requires larger molecular fractions
than used for the UV absorbers ($f_{\rm H2} \approx 0.01$).

This initial survey opens up considerable opportunities for future 
studies of the \Htwo--cirrus correlation.  Higher resolution infrared 
maps with more accurate temperature corrections, 
such as those obtainable with the {\em Spitzer Space Telescope}, would 
greatly improve the effectiveness of this method.  A comparison with
H~I 21-cm maps (e.g., Lockman \& Condon 2005) could delineate the gaseous 
structures associated with the cirrus and measure the dust-to-gas
ratios in these diffuse clouds.  Expanding the UV sample to 
include more sight lines to background AGN would improve the statistics 
of such a small sample.  The 45 AGN in the survey by Gillmon
\etal\ (2005) were a subset of the 219 {\em FUSE} targets selected in Wakker
\etal\ (2003) as candidates for the analysis of Galactic O\,VI.  The next 
50 brightest targets have an average 
flux of $6\times10^{-14}$~erg~cm$^{-2}$ s$^{-1}$~\AA$^{-1}$.  
To achieve a S/N of 3 per pixel with {\em FUSE} would require 
approximately 20 ksec per AGN, for a program total of 1000 ksec.  
An ultraviolet telescope with sensitivity greater than FUSE is 
probably necessary for feasible exposure times in a $50^{+}$ target survey.  

Another promising avenue for future exploration is to compare the
cirrus maps and \Htwo\ absorption lines with other tracers, such as
H~I, CO and $\gamma$-ray emission.  On the basis of such comparisons,
Grenier, Casandjian, \& Terrier (2005) suggest that many interstellar 
clouds in the solar neighborhood
have extensive dark regions that bridge the dense cloud cores to 
atomic phases.  These details are beyond the scope of our current paper. 
Perhaps the most direct way to investigate the connection between diffuse 
\Htwo\ and IR cirrus would be to map H$_2$ in UV or IR fluorescent emission.  
This method would provide the morphological information lacking in  
``pencil-beam" UV-absorption sight lines.  There is currently no 
high-resolution experiment that can map diffuse H$_2$ emission, either in 
the mid-infrared (28 and 17 $\mu$m) or the far-ultraviolet (1000--1100 \AA).  
Intriguing results for \Htwo\ ultraviolet fluorescent emission at
$10'$ resolution may soon be available from the {\it Spectroscopy of Plasma 
Evolution from Astrophysical Radiation} (SPEAR) Mission (Edelstein \etal\ 2003). 
With appropriate IR and UV telescopes, these methods could 
help map the gaseous Galactic halo. 

\acknowledgments{} 

We thank Ken Sembach and Jay Lockman for useful discussions. 
This work was based in part on data obtained for the Guaranteed 
Time Team team by the NASA-CNES-CSA \FUSE\ mission operated by 
the Johns Hopkins University.  Financial support to U.S. 
participants has been provided by NASA contract NAS5-32985.
The Colorado group also received \FUSE\ support from
NASA grant NAG5-10948 for studies of interstellar \Htwo.

\newpage


\clearpage


\begin{deluxetable}{lccccc} 
\tablecolumns{6}
\tablenum{1}
\tablewidth{0pt}
\tablecaption{Additional FUSE Targets Behind Regions of Low Cirrus}
\tablehead{ 
\colhead{Target} & \colhead{$l$} & \colhead{$b$} &
       \colhead{$D_{100}^{(T)}$} & \colhead{S/N} & \colhead{N$_{\rm H2}$ } \\
\colhead{}  & \colhead{(deg)} & \colhead{(deg)} & \colhead{(MJy~sr$^{-1}$)} & 
       \colhead{(pixel$^{-1}$)} & \colhead{(cm$^{-2}$)} 
} 
\startdata
MS0354-3650  & 238.87 & $-50.35$ & 0.37 &  3 & $\leq 14.40$  \\
HAR03        & 152.41 &   52.87  & 0.38 &  2 & $\leq 14.52$  \\
UGC5720      & 156.20 &   52.80  & 0.41 &  3 & $18.79 \pm 0.05$ \\
NGC5447      & 102.82 &   59.83  & 0.44 &  2 & $\leq 14.52$  \\
M101D        & 102.67 &   59.75  & 0.46 &  2 & $\leq 14.50$   
\enddata
\end{deluxetable}


\begin{deluxetable}{lccc}
\tablecolumns{4}
\tablenum{2}
\tablewidth{0pt}
\tablecaption{Selected Dust-to-Gas Ratios\tablenotemark{a} }
\tablehead{
\colhead{Name} & \colhead{$D_{100}^{(T)}$} & \colhead{log~N$_{\rm HI}$} & 
   \colhead{$D_{100}^{(T)}$/N$_{\rm HI}$} \\ 
\colhead{} & \colhead{(MJy~sr$^{-1}$)} & \colhead{(cm$^{-2}$)} & 
   \colhead{($10^{-20}$~MJy~sr$^{-1}$~cm$^2$)}
}
\startdata
{3C 249.1}     &  1.98 &  20.14 &  1.43  \\
{3C 273}       &  1.13 &  19.42 &  4.29  \\
{H 1821+643}   &  2.34 &  20.34 &  1.07  \\
{HS 0624+6907} &  5.29 &  20.62 &  1.27  \\
{MRC 2251-178} &  2.14 &  20.12 &  1.62  \\
{Mrk 205}      &  2.28 &  20.25 &  1.28  \\
{Mrk 421}      &  0.83 &  19.73 &  1.54  \\
{PG 0804+761}  &  1.92 &  20.41 &  0.74  \\
{PG 0844+349}  &  2.00 &  20.24 &  1.15  \\
{PG 0953+414}  &  0.66 &  20.00 &  0.66  \\
{PG 1116+215}  &  1.23 &  19.83 &  1.82  \\
{PG 1211+143}  &  1.86 &  20.33 &  0.87  \\
{PG 1259+593}  &  0.44 &  19.67 &  0.95  \\
{PG 1302-102}  &  2.36 &  20.22 &  1.42  \\
{PKS 0405-12}  &  3.22 &  20.28 &  1.69  \\
{PKS 2155-304} &  1.19 &  20.06 &  1.03  \\
\enddata

\tablenotetext{a}{Ratio, $D_{100}^{(T)}$/\NHI, of 100 $\mu$m cirrus
emission from SFD98 (MJy~sr$^{-1}$) to H~I column density (\cd) 
toward 16 AGN sight lines in our survey observed
in 21 cm by Lockman \& Savage (1995). The mean and standard 
deviation, excluding the anomalous 3C~273 sight line, are 
$(1.43 \pm 0.41) \times 10^{-20}$ (MJy~sr$^{-1}$~cm$^2$), 
in excellent agreement with Boulanger \etal\ (1985). }  

\end{deluxetable}

\clearpage


\begin{figure}
\epsscale{1.0}
\plottwo{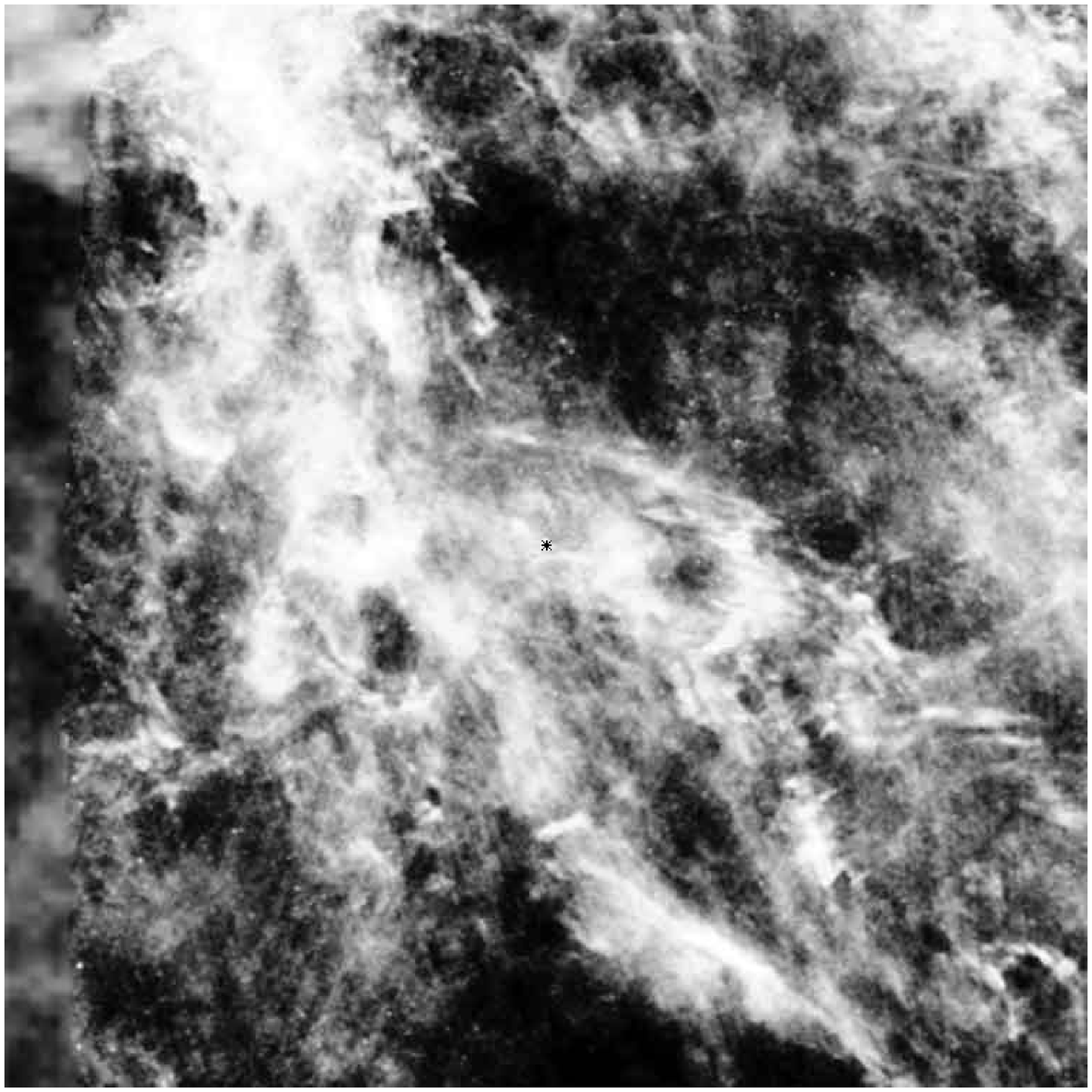}{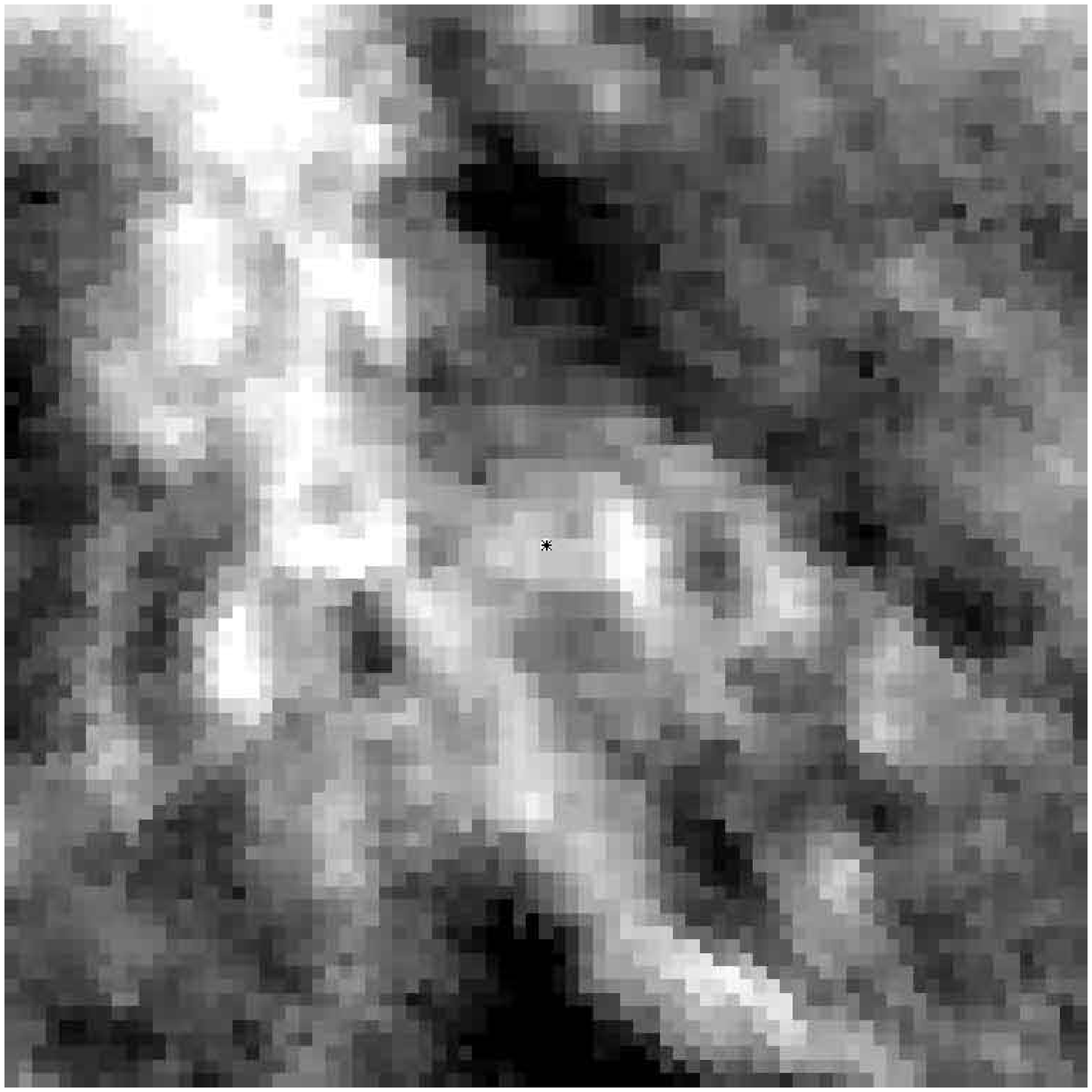} 
\caption{\em Left panel\normalfont: \IRAS\ temperature-corrected 
  100~$\mu$m flux map (SFD98).
  \it Right panel\normalfont: 21~cm emission map from Leiden-Dwingeloo 
  survey (Hartmann \& Burton 1997). Both maps show the same 
  $30^{\circ}\times30^{\circ}$ field centered at 
  $(l,b)=(267.55^{\circ},74.32^{\circ})$, the location of 
  the AGN, PG\,1211+143, one of the sight lines studied by Gillmon
  \etal\ (2005) and Tumlinson \etal\ (2005). } 
\end{figure}


\begin{figure}
\epsscale{1.0}
\plotone{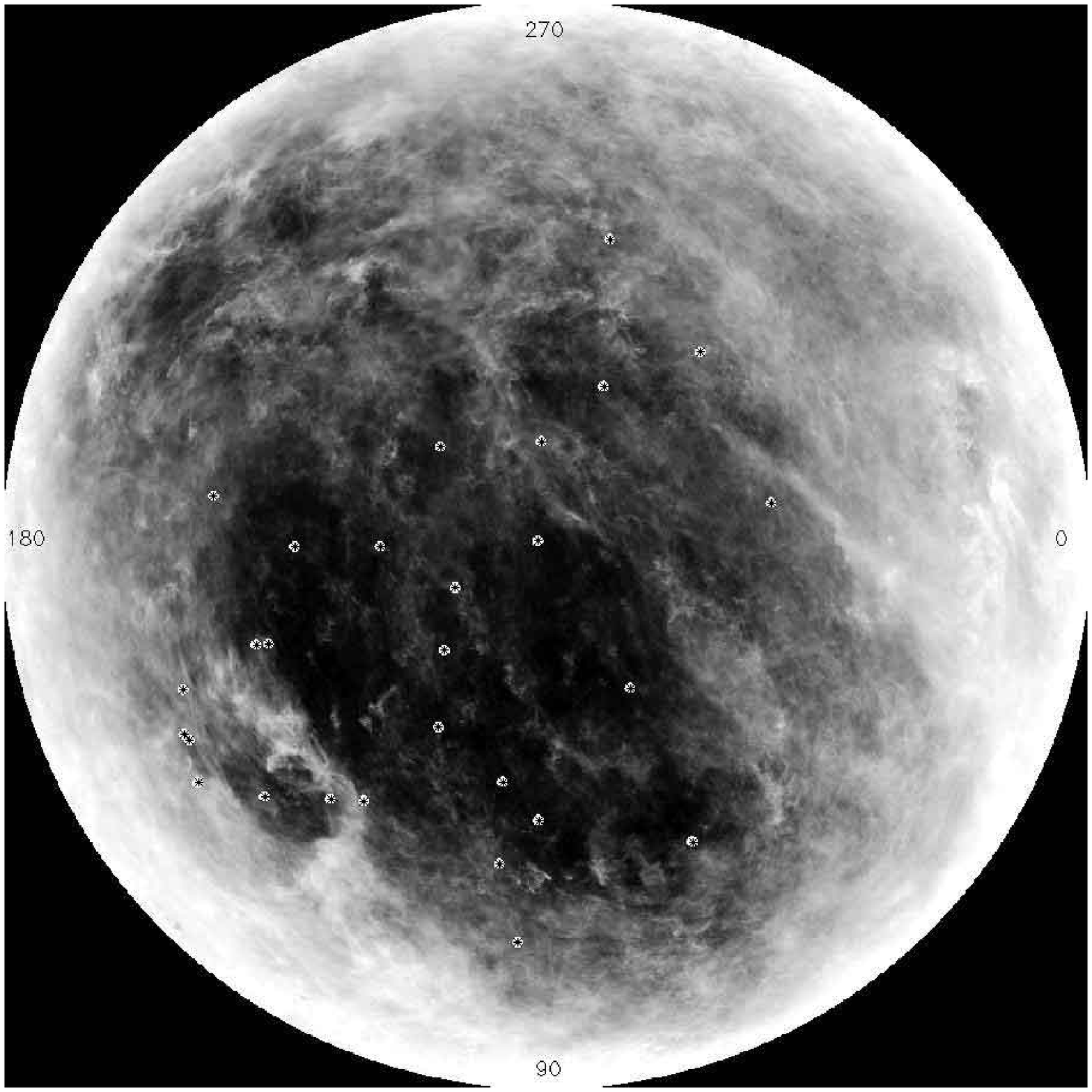} 
\caption{\IRAS\ temperature-corrected 100~$\mu$m flux map (SFD98) 
   for the northern Galactic hemisphere, centered on 
   $b=90^{\circ}$ with $b=0^{\circ}$ around the edge.  Numbers 0, 90, 180, 
   and 270 indicate Galactic longitude, and asterisks mark locations of 
   28 AGN sight lines (Gillmon \etal\ 2005). } 
 
\end{figure}


\begin{figure}
\epsscale{1.0}
\plotone{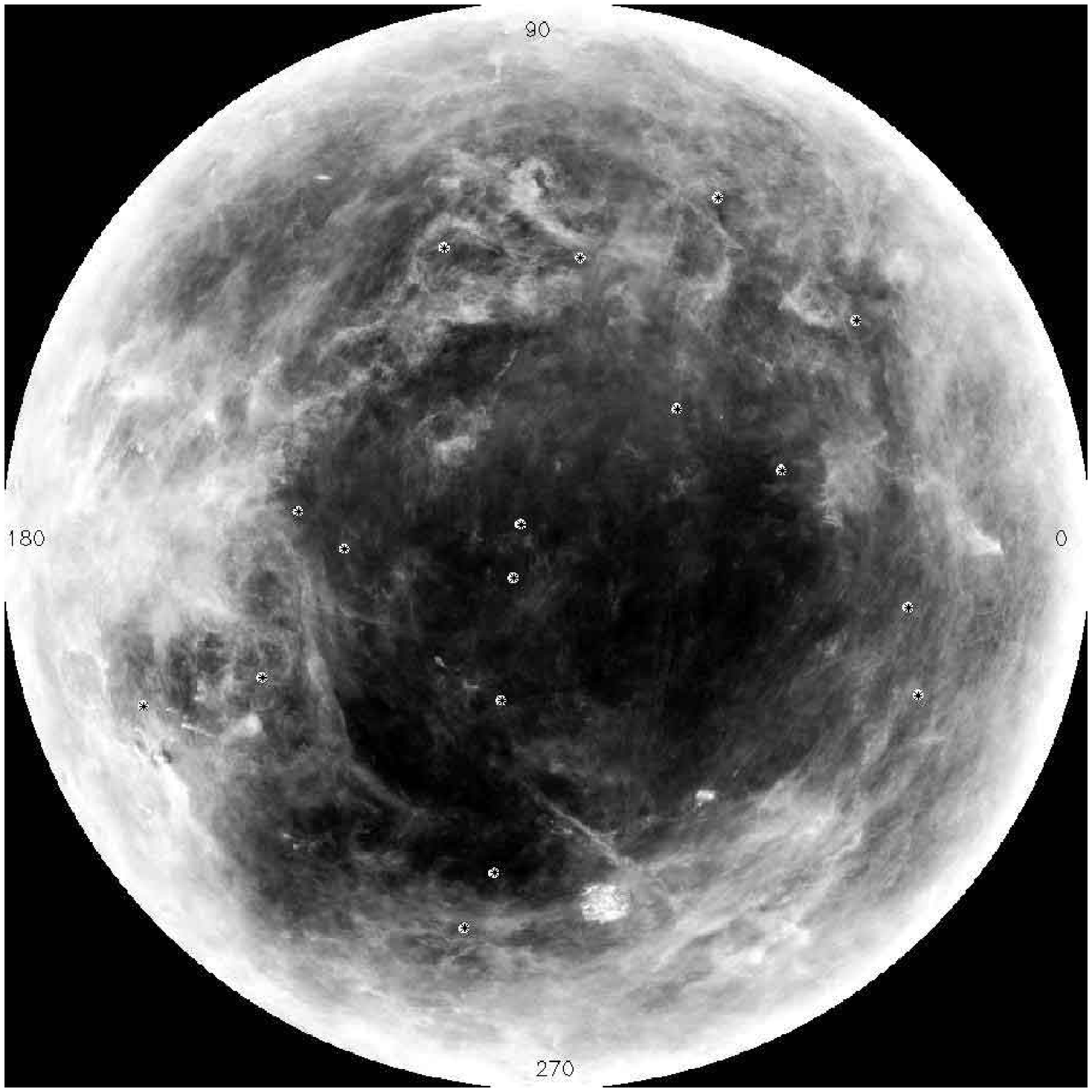} 
 \caption{\IRAS\ temperature-corrected 100~$\mu$m flux map 
   (SFD98) for the southern Galactic hemisphere, centered on 
   $b=-90^{\circ}$ with $b=0^{\circ}$ around the edge.  
   Numbers 0, 90, 180, and 270 indicate Galactic longitude, and   
   asterisks mark locations of 17 AGN sight lines (Gillmon
   \etal\ 2005).  }  
\end{figure}


\begin{figure}
\epsscale{1.0}
\plotone{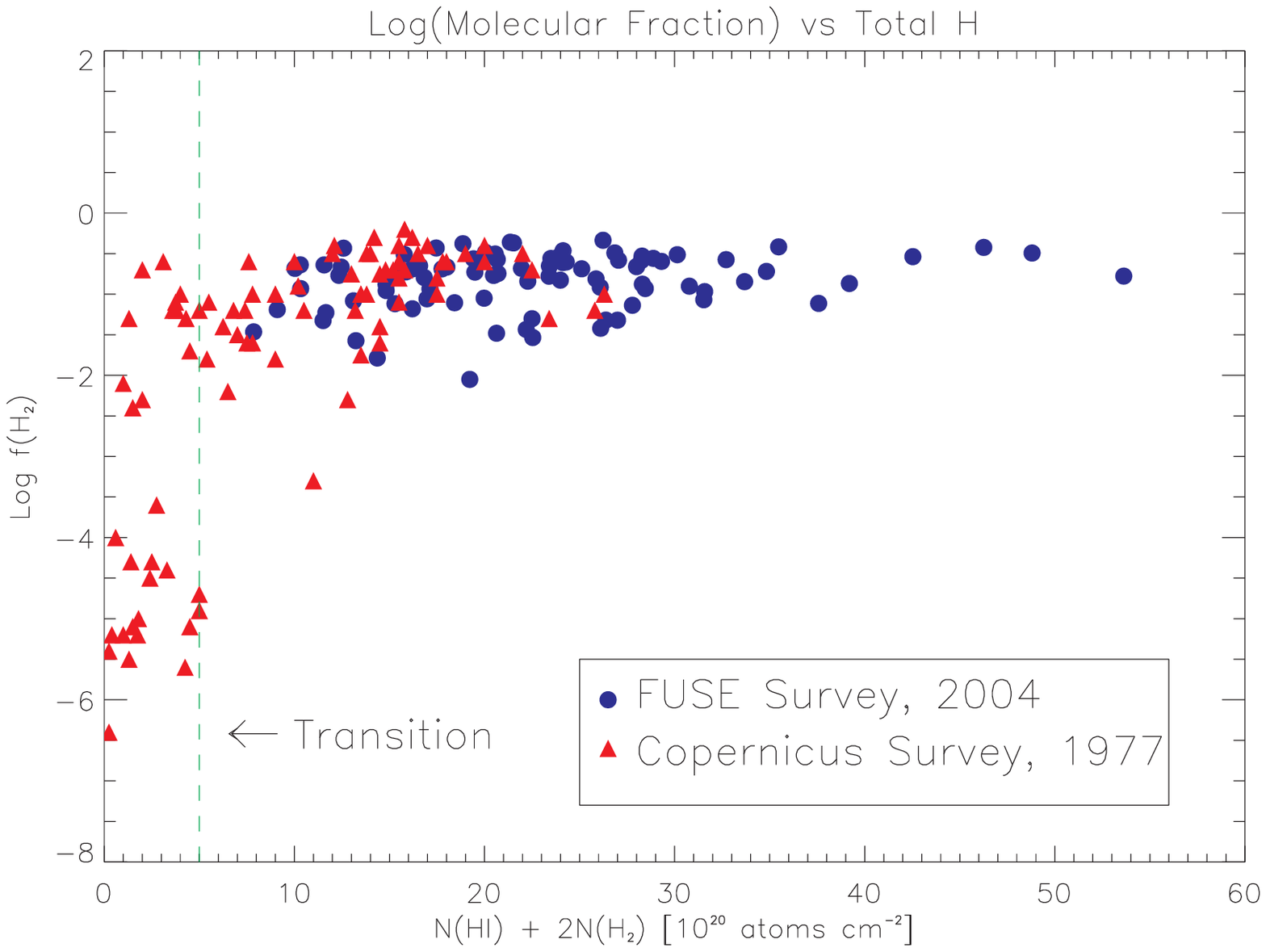} 
\caption{Results from {\it Copernicus} and \FUSE\ surveys of 
  H$_2$ showing the self-shielding transition to higher $f_{\rm{H2}}$ at 
  N$_{\rm{H}}\ge5\times10^{20}$~cm$^{-2}$ (Savage \etal\ 1977; 
  Shull \etal\ 2005).}
\end{figure}


\begin{figure}
\epsscale{1.0}
\plotone{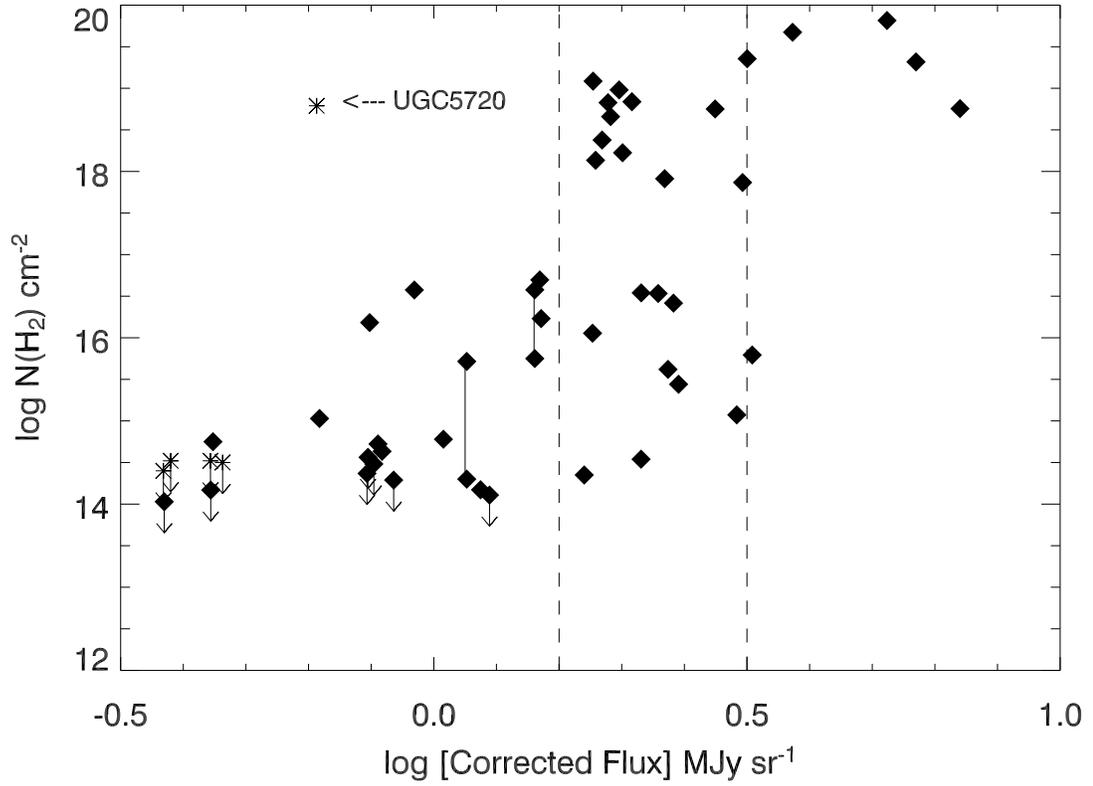} 
\caption{Column density of molecular hydrogen, \NHtwo, toward the 
   45 AGN in the \FUSE\ survey (Gillmon \etal\ 2005, blue diamonds) 
   vs.\ temperature-corrected 100~$\mu$m intensity, $D_{100}^{(T)}$, 
   from the \IRAS\ maps (SFD98).  Arrows indicate upper limits on \NHtwo.  
   Asterisks show five additional sight lines analyzed for this work.
   The self-shielding transition of H$_2$ is visible between 
   log~$D_{100}^{(T)} \approx (0.2-0.5)$ [$D_{100}^{(T)} = 1.5-3.0$ 
   MJy~sr$^{-1}$], indicating that a significant amount of the detected 
   \Htwo\ is in the infrared cirrus. The molecular fractions,
   $f_{\rm H2}$, range from 1--30\% above this transition. } 
\end{figure}


\begin{figure}
\epsscale{1.0}
\plotone{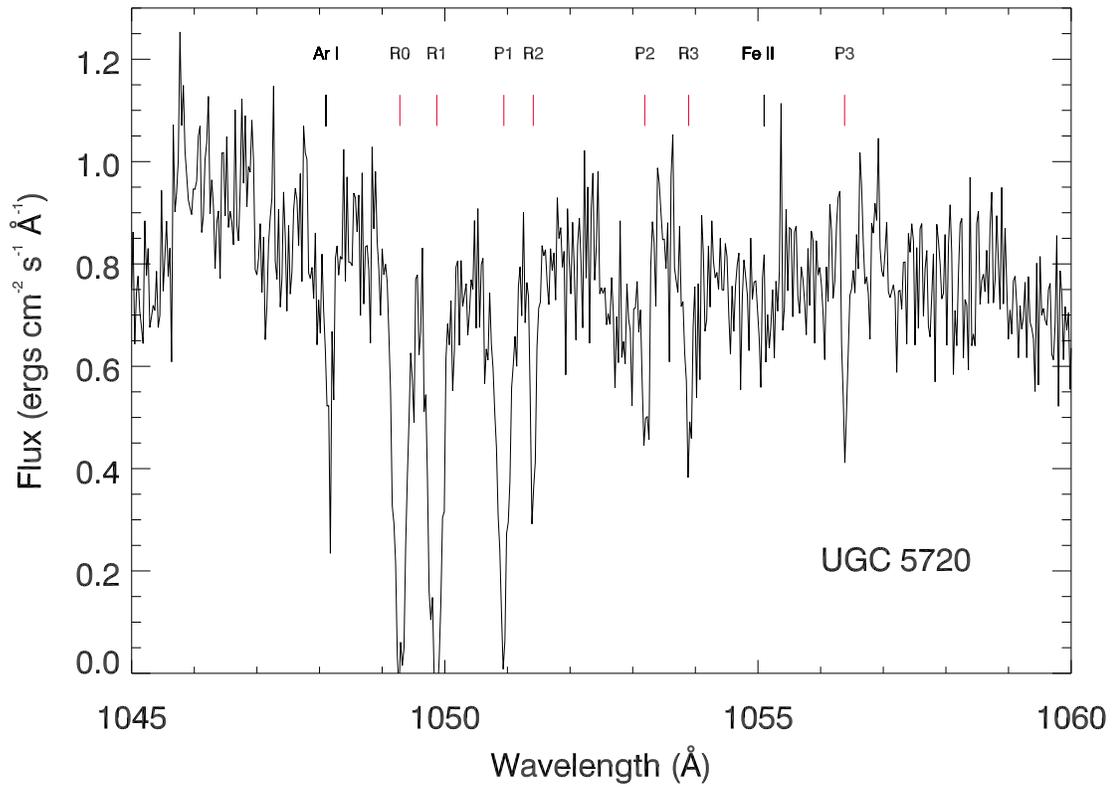} 
\caption{Portion of the \FUSE\ spectrum showing the Lyman (4-0) band 
   of H$_2$ toward UGC~5720.  This sightline has log~\NHtwo\ 
   $= 18.79 \pm 0.05$.}
\end{figure}


\begin{figure}
\epsscale{1.0}
\plotone{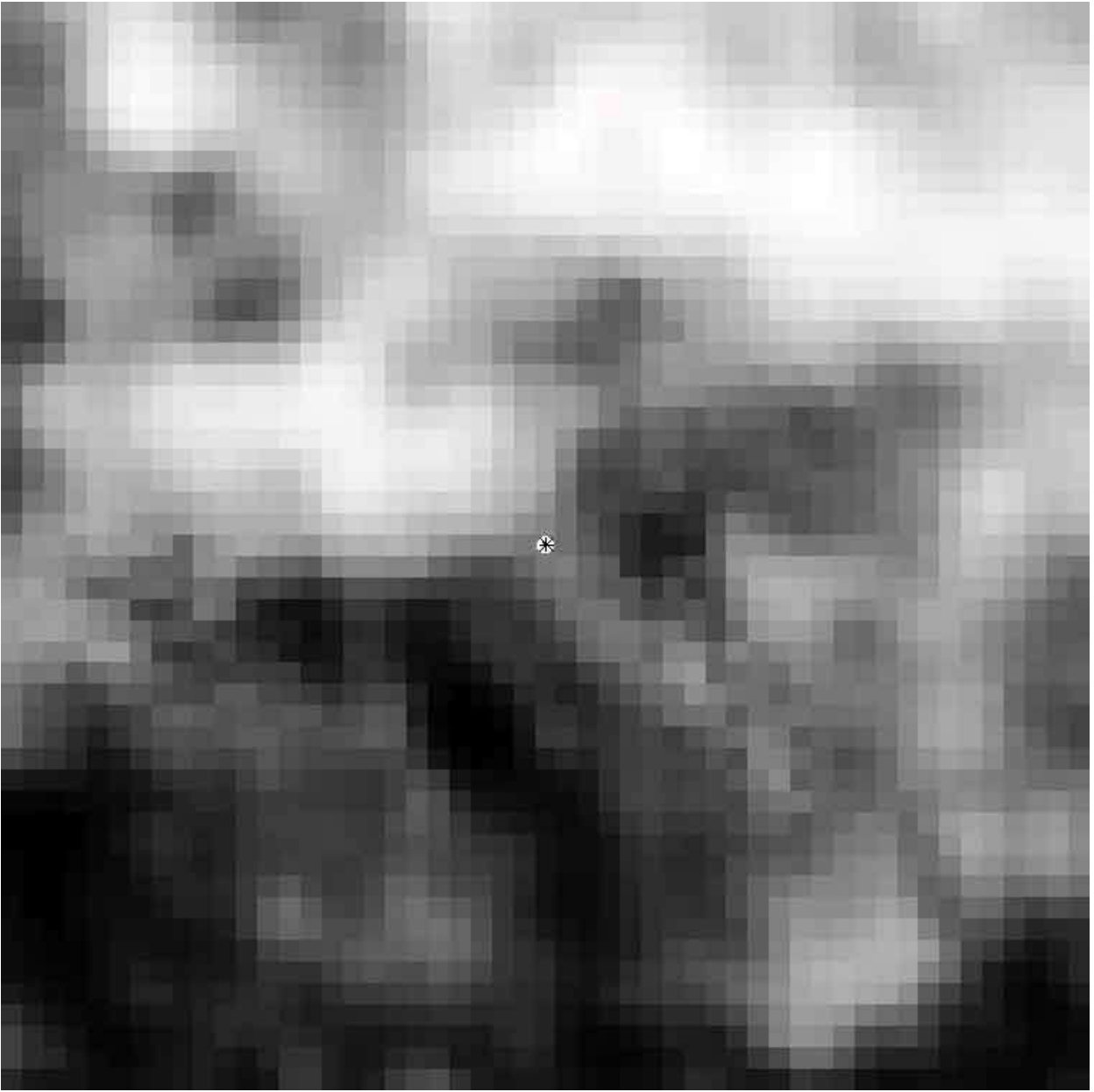} 
\caption{\IRAS\ temperature-corrected 100~$\mu$m map (\S~2.2)  
    in a $2^{\circ}\times2^{\circ}$ field centered at 
    $(l,b)=(156.20^{\circ},52.80^{\circ})$.  
    Asterisk marks location of UGC\,5720, the anomalous sight line with 
    low temperature-corrected flux and high H$_2$ column density 
    discussed in \S~3.2.}
\end{figure}


\begin{figure}
\epsscale{1.0}
\plotone{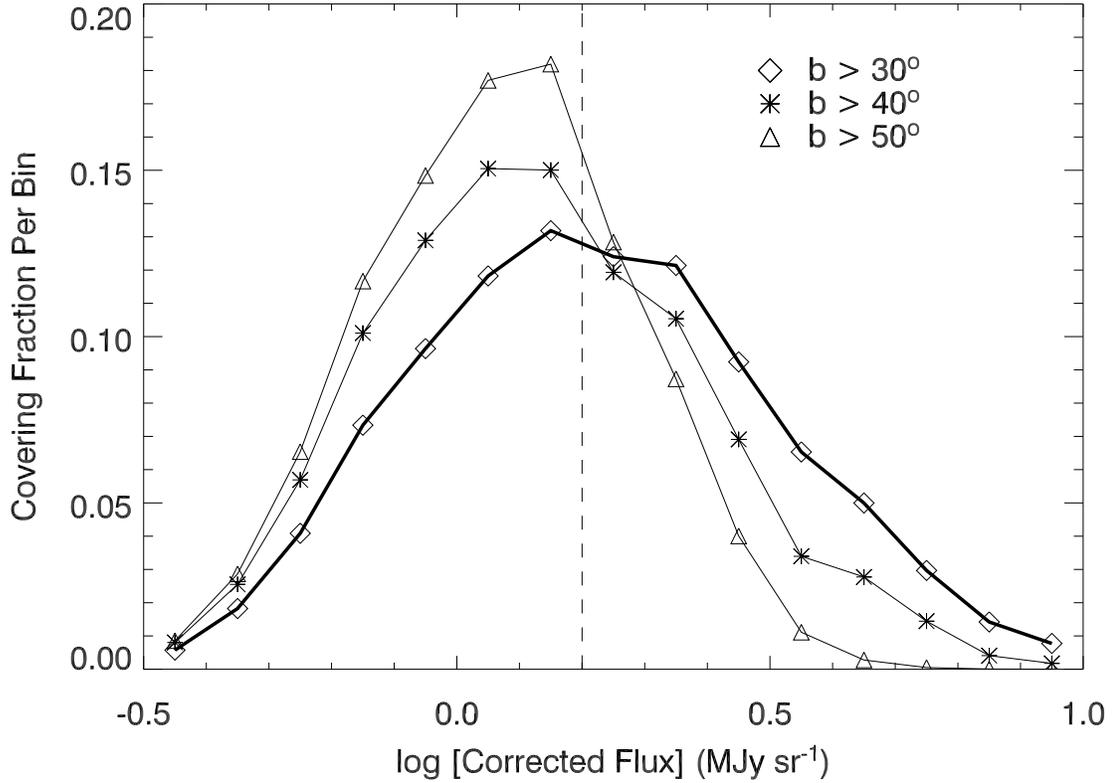} 
\caption{Area covering fraction of 100~$\mu$m cirrus in the
    northern Galactic hemisphere (SFD98) vs. cirrus intensity,
    log~$D_{100}^{(T)}$ (MJy~sr$^{-1}$).  The three curves show the 
    differential distribution, in logarithmic bins of width 0.1 
    (MJy~sr$^{-1}$), for the sky above latitudes $b = 30^{\circ}$, 
    $40^{\circ}$, and $50^{\circ}$. Vertical dashed line shows
    the inferred H~I/\Htwo\ transition (see Figure 5). }   
\end{figure}

\end{document}